\documentclass[apj]{emulateapj}
\usepackage{amsmath}
\usepackage{apjfonts}

\newcommand{\DDir}{\relax{D\kern-.7em{/}}}

\newcommand{\X}{\times}

\newcommand{\be}{\begin{equation}}
\newcommand{\ee}{\end{equation}}
\newcommand{\bea}{\begin{equation*}}
\newcommand{\eea}{\end{equation*}}

\newcommand{\nin}{\relax{\in\kern-.8em{/}}}

\newcommand{\bt}{\beta}

\newcommand{\de}{\delta}

\newcommand{\vep}{\varepsilon}
\newcommand{\ep}{\epsilon}
\newcommand{\zt}{\zeta}

\newcommand{\cm}{\mbox{ cm}}

\newcommand{\se}{\mbox{ s}}
\newcommand{\yr}{\mbox{ yr}}

\newcommand{\erg}{\mbox{ erg}}

\newcommand{\km}{\mbox{ km}}
\newcommand{\pc}{\mbox{ pc}}

\newcommand{\eV}{\mbox{ eV}}

\newcommand{\GeV}{\mbox{ GeV}}
\newcommand{\TeV}{\mbox{ TeV}}

\newcommand{\muG}{\mbox{ $\mu$G}}

\newcommand{\vv}{\textrm{v}}

\begin{document}

\title{Cosmic rays from trans-relativistic supernovae}

\author{Ran Budnik\altaffilmark{1}, Boaz Katz\altaffilmark{1}, Andrew MacFadyen\altaffilmark{2,3} and Eli Waxman\altaffilmark{1}}

\altaffiltext{1}{Physics Faculty, Weizmann Institute, Rehovot 76100, Israel}
\altaffiltext{2}{Institute for Advanced Study, Princeton, NJ 08540, USA}
\altaffiltext{3}{Physics Department, New York University, NY, NY 10003, USA}

\begin{abstract}
We derive constraints that must be satisfied by the sources of $\sim10^{15}$ to $\sim10^{18}$~eV cosmic rays, under the assumption that the sources are Galactic. We show that while these constraints are not satisfied by ordinary supernovae (SNe), which are believed to be the sources of  $\lesssim10^{15}$~eV cosmic rays, they may be satisfied by the recently discovered class of trans-relativistic supernovae (TRSNe), which were observed in association with gamma-ray bursts. We define TRSNe as SNe that deposit a large fraction, $f_R>10^{-2}$, of their kinetic energy in mildly relativistic, $\gamma\beta>1$, ejecta. The high velocity ejecta enable particle acceleration to $\sim10^{18}$~eV, and the large value of $f_R$ (compared to $f_R\sim10^{-7}$ for ordinary SNe) ensures that if TRSNe produce the observed $\sim10^{18}$~eV cosmic ray flux, they do not overproduce the flux at lower energies. This, combined with the estimated rate and energy production of TRSNe, imply that Galactic TRSNe may be the sources of cosmic rays with energies up to $\sim 10^{18}\eV$.
\end{abstract}


\section{Introduction}
\label{sec:Introduction}

The  spectrum of cosmic rays (CRs) extends from $\sim10^9$~eV to $\sim10^{20}$~eV \citep[see][for reviews]{BE87,Axford94,NaganoWatson}. While there is a wide consensus that the sources of CRs of energy $\lesssim10^{15}$~eV are Galactic supernovae (SNe), and that the sources of CRs above $10^{19}$~eV are extra-Galactic \citep[possibly GRBs or AGN, e.g.][for a recent review]{Waxman04c}, there is no consensus regarding the sources of cosmic rays in the energy range of $\sim10^{15}$~eV to $\sim10^{18}$~eV.

The CR spectrum steepens at $\sim10^{15}$~eV, the "knee," and then extends smoothly to $\sim10^{18}$~eV. These
characteristics disfavor a transition from Galactic to extra-Galactic sources in this energy range, since for such a transition one would naturally expect either a flattening of the spectrum or a marked drop of the flux at the transition energy. In fact, these characteristics suggest that the same type of sources, or the same class of sources, produces the CRs over the entire energy range of $\sim10^9$~eV to $\sim10^{18}$~eV. Indeed, it has been suggested that Galactic SNe are the sources of all CRs in this energy range \citep[e.g.][]{BellLucek}.

Over the past several years a new type of supernova explosions has been discovered, to which we will refer as "trans-relativistic supernovae" (TRSNe). We define TRSNe as supernovae which, unlike ordinary SNe, deposit a significant fraction, $f_R>10^{-2}$, of their kinetic energy in mildly relativistic, $\gamma\beta>1$, ejecta (the corresponding fraction for ordinary SNe is typically $\le 10^{-7}$; see \S~\ref{sec:TRSNe} for a detailed discussion). Such TRSNe have been discovered recently in association with gamma-ray bursts (GRBs). Hereafter we will use the term SNe for ordinary supernovae, which deposit only a small fraction of their kinetic energy in relativistic ejecta, and the term TRSNe for supernova explosions with $f_R>10^{-2}$.

Various terms have been used in the literature to describe supernovae associated with GRBs: GRB-SNe, "hypernovae", broad-line type Ic SNe, and others. We have chosen here to use the term "trans-relativistic supernovae" for several reasons. First, it captures the main physical property that characterizes supernovae that may produce the observed $\sim10^{15}$~eV to $\sim10^{18}$~eV cosmic-ray flux, namely $f_R>10^{-2}$. Second, supernovae associated with GRBs may not all be similar to each other. SN2006aj, for example, is qualitatively different from the other GRB associated supernovae. Although there is evidence for this explosion to be characterized by $f_R>10^{-2}$, its total kinetic energy is similar to that of ordinary SNe, $\sim10^{51}$~erg, while the other GRB associated supernovae appear to be ten times more energetic. Thus, while SN2006aj does not fit into the "hypernovae" (i.e. highly energetic) category, it does fit into the TRSNe category and should be considered as a potential source of $\sim10^{15}$~eV to $\sim10^{18}$~eV cosmic-rays. Finally, it should be kept in mind that not all TRSNe are necessarily associated with GRBs.

We show in this letter that Galactic TRSNe may be the sources of $\sim10^{15}$~eV to $\sim10^{18}$~eV CRs. We first derive in \S~\ref{sec:physical_constraints} constraints, which apply to both SNe and TRSNe, that must be satisfied by Galactic sources of $\sim10^{15}$~eV to $\sim10^{18}$~eV. We then show in \S~\ref{sec:Application} that while these constraints are not satisfied by SNe, they may be satisfied by TRSNe. Our results are summarized and their implications are discussed in \S~\ref{sec:conclusions}.

\section{Constraints on Galactic sources of CRs}
\label{sec:physical_constraints}

We consider three constraints which must be satisfied by the candidate sources of CRs at high energy, $\vep\sim10^{18}$~eV. First we consider in \S~\ref{sec:Rates} constraints on the rate of occurrence of the sources in the Galaxy and on their energetics. The constraints derived in \S~\ref{sec:Rates} are applicable to any (Galactic) candidate source. Next, we consider in \S~\ref{subsec:Maximum-energy} and \S~\ref{sec:ejecta} constraints that are applicable to SNe and TRSNe. These sources are expected to accelerate particles to high energies through the collisionless shocks which they drive into the plasma surrounding the exploding star. In \S~\ref{subsec:Maximum-energy} we derive constraints on the velocity of the shell ejected by the explosion, and in \S~\ref{sec:ejecta} we derive constraints on the structure of the ejecta.

\subsection{Rates and energetics}
\label{sec:Rates}

In order for a certain type of sources to be the main contributor to the observed CR flux at some energy $\vep_0$, the sources must satisfy the following conditions:
\begin{enumerate}
\item The energy production rate at $\vep_0$ should be sufficient to explain the observed flux, and
\item The Galactic event rate, that is the rate of occurrence of the sources in the Galaxy, should not be much lower than one event per Galactic confinement time of CRs with energy $\vep_0$.
\end{enumerate}
Consider a source with a Galactic occurrence rate of $\dot{N}_{\rm s}$ and an energy per event of $E_{\rm s}$. $E_{\rm s}$ denotes the energy available for the acceleration of $10^{18}\eV$ CRs, rather than the total energy released. In SN explosions, for example, the energy available for CR acceleration is the energy deposited in the shock wave driven by the ejecta. Since the maximum energy to which particles may be accelerated increases with shock velocity (see \S~\ref{subsec:Maximum-energy}), the energy available for acceleration of particles to $10^{18}\eV$ is the energy carried by the fastest (outermost) part of the ejecta, which drives, at early times, a sufficiently fast shock. This energy may be significantly smaller than the total kinetic energy of the (slower) ejecta.
We assume that the source produces an energy spectrum of accelerated particles given by
\begin{equation}\label{eq:sourcespectrum}
\vep^2\frac{dN}{d\vep}=\zt_{\rm s}(\vep)E_{\rm s}.
\end{equation}
Here $\zt_{\rm s}(\vep)$ is the fraction of $E_{\rm s}$ that is deposited in CRs within a logarithmic energy interval around $\vep$. The resulting energy density of CRs per unit logarithmic particle energy is
\begin{equation}\label{eq:densitygeneral}
\vep^2n(\vep)=\frac{\zt_{\rm s}(\vep)\dot N_{\rm s}\tau_{c}(\vep)E_{\rm s}}{V_G},
\end{equation}
where $n(\vep)$ is the number density per unit CR energy, $\tau_{c}$ is the confinement time of the particles in the Galaxy, and $V_G$ is the effective volume of the Galaxy in which the particles are confined. We assume that this effective volume is energy independent.

To reduce uncertainties we compare the resulting densities with those of low
energy, $10^9$~eV, CRs, assuming that in this energy range the CRs are
accelerated predominantly by SNe.
Using Eq.~(\ref{eq:densitygeneral}), the ratio of CR number densities (per unit CR energy) at low and high energy, $n_{l,h}=n(\vep=\vep_{l,h})$, is
\begin{equation}\label{eq:energeticconst1}
\frac{\vep_h^2n_{h}}{\vep^2_ln_{l}}\approx\frac{\dot N_{\rm s}}{\dot N_{\rm SN}}\frac{\tau_{c}(\vep_{h})}{\tau_{c}(\vep_{l})}\frac{E_{\rm s}}{E_{\rm SN}}\frac{\zeta_{\rm s}(\vep_h)}{\zeta_{\rm SN}(\vep_l)}, \end{equation}
where $E_{\rm SN}$ is the characteristic kinetic energy of a SN explosion.
For $\vep_h=10^{18}\eV$ and $\vep_l=10^9\eV$ the observed ratio is \citep[e.g. ][and references within]{NaganoWatson}
\begin{equation}\label{eq:obsfluxratio}
\frac{\vep_h^2n_{h}}{\vep^2_ln_{l}}
\approx(10^{15}/10^{9})^{-0.7}(10^{18}/10^{15})^{-1.0}\simeq10^{-7}.
\end{equation}

Using Eqs. (\ref{eq:energeticconst1}) and (\ref{eq:obsfluxratio}), we find that the sources of CRs at $\vep=\vep_h=10^{18}\eV$ must satisfy
\begin{equation}\label{eq:num_of_sources1}
\frac{N_{\rm s}}{N_{\rm SN}}\approx 10^{-7}\zt^{-1}\left(\frac{E_{\rm s}}{E_{\rm SN}}\right)^{-1},
\end{equation}
where $N_{\rm s}=\dot N_{\rm s}\tau_{c}(\vep_h)$ and $N_{\rm SN}=\dot N_{\rm SN}\tau_{c}(\vep_l)$ are the average numbers of events that contribute to the observed flux at any given time at energies $\vep_h=10^{18}\eV$ and $\vep_l=10^{9}\eV$ respectively, and $\zt\equiv\zeta_{\rm s}(10^{18}\eV)/\zeta_{\rm SN}(10^9\eV)$. Using the confinement time of 1~GeV protons in the Galaxy, $t_{c}(10^9\eV)\approx10^{7.5}\yr$ \citep{Webber98, Yanasak01}, Eq.~(\ref{eq:num_of_sources1}) may be written as
\begin{equation}\label{eq:num_of_sources2}
N_{\rm s}\approx 0.03\zt^{-1}\left(\frac{E_{\rm s}}{E_{\rm SN}}\right)^{-1}\dot N_{SN,-2},
\end{equation}
where  $\dot N_{\rm SN}=0.01\dot N_{SN,-2}\yr^{-1}$. The requirement that the event rate should exceed one event per Galactic confinement time, $N_{\rm s}\gtrsim1$, implies
\begin{equation}\label{eq:jointconst}
\zeta\frac{E_{\rm s}}{E_{\rm SN}}<0.03\dot N_{SN,-2}.
\end{equation}

The confinement time of CRs of energy $\sim10^{18}\eV$ is likely to be significantly larger than the light crossing time of the Galaxy, $\sim10^{4}\mathrm{yr}$, due to the following argument. The observed energy evolution of the depth of maximum, $X_{\rm max}$, of extreme air showers suggest that composition of CRs at $\sim10^{18}\mathrm{eV}$ is dominated by nuclei with moderate atomic number $Z$ \citep{Gaisser93,NaganoWatson}. Consider therefore the Larmor radius of a CR particle with atomic number $Z=10Z_{1}$ and an energy $\vep=10^{18}\vep_{18}\mathrm{eV}$, \[ R_{L}\approx 40 B_{-5.5}^{-1}\vep_{18}Z_{1}^{-1}\pc,\] where $B=3B_{-5.5}\mu G$ is the Galactic magnetic field. Since the Larmor radius of such particles is much smaller than the thickness of the CR disk, $\sim1$~kpc, the propagation of $\vep=10^{18}\mathrm{eV}$ CRs is expected to be strongly affected by the Galactic magnetic field. The transition to non confined particles, for which the confinement time is comparable to the light crossing time, should then occur at $\gtrsim10^{19} Z_1\eV$. Using therefore $\tau_{c}(10^{18}\eV)=\tau_{4.5}10^{4.5}\mathrm{yr}$, Eqs. (\ref{eq:energeticconst1}) and (\ref{eq:obsfluxratio}) give
\begin{equation}\label{eq:energeticconst2}
\frac{\dot N_{\rm s}}{\dot N_{\rm SN}}=10^{-4}\tau_{4.5}^{-1}\left(\frac{E_{\rm s}}{E_{\rm SN}}\right)^{-1}\zt^{-1}.\end{equation}

The constraint that the Galactic event rate be larger than one event per confinement time can be written as:
\begin{equation}\label{eq:energeticconst3}
\frac{\dot N_{\rm s}}{\dot N_{\rm SN}}>3\times10^{-3}\tau_{4.5}^{-1}\dot N_{SN,-2}^{-1}.\end{equation}
Eqs.~(\ref{eq:jointconst}),~(\ref{eq:energeticconst2}) and (\ref{eq:energeticconst3}) (any one of these can be derived from the other two) describe the constraints on the occurrence rate and energetics of the sources of $\sim10^{18}$~eV CRs.

A note is in order regarding the value of the parameter $\zt\equiv\zt_s(10^{18}\eV)/\zt_{\rm SN}(10^9)\eV$. In the following, we consider sources that accelerate particles through collisionless shocks in the surrounding medium driven by fast ejecta. Note that although a complete theory, based on first principles, of particle acceleration in collisionless shocks is not available, there are strong theoretical arguments and observational evidence indicating that for both non-relativistic and relativistic shocks a considerable fraction of the post shock energy can be converted to relativistic particles with a spectrum that follows $\vep^2 dN/d\vep\approx\vep^0$ \citep[see, e.g.][for reviews]{BE87,Axford94,Waxman06}.
If both SNe and the sources of $10^{18}$~eV CRs deposit a significant fraction of their energy in accelerated particles with a spectrum $dN/d\vep\propto \vep^{-2}$, the value of $\zt$ would be of order unity.

\subsection{Maximum energy
\label{subsec:Maximum-energy}}

We assume that the particles are accelerated in the vicinity of a collisionless shock wave with (a time dependent) velocity $\vv_S=\bt_S c$ that is driven into the surrounding medium by an ejecta of mass $M_{\rm ej}$, initial velocity $\vv_{\rm ej}=\bt_{\rm ej}c$ and kinetic energy $E_{\rm ej}=(\gamma_{\rm ej}-1)M_{\rm ej}c^2$. Here $\gamma_{\rm ej}=(1-\bt_{\rm ej}^2)^{1/2}$ is the ejecta's Lorenz factor. We consider two options for the density distribution of the surrounding medium: A homogeneous inter-stellar medium (ISM) or a progenitor wind with a density profile $\rho\propto r^{-2}$.

At any given time, the maximum energy to which particles can be accelerated is given by
\begin{equation}\label{eq:maxegeneral} \vep_{\rm max}\sim Ze\bt_SB_dr_S \end{equation}
\citep[e.g.][for a recent review]{Waxman04c}. Here, $r_S$ is the radius of the shock and $B_d$ is the minimum of the magnetic field in the pre-shock (upstream) and post-shock (downstream) plasma, measured in downstream frame. We should stress that the energy given by Eq.~(\ref{eq:maxegeneral}) is an upper limit, and that the maximum energies attained are likely to be lower. For example, if particles are accelerated via Fermi's diffusive shock acceleration, the maximal energy would be a few times lower for Bohm limit diffusion, and much lower for slower diffusion.

There is growing evidence that the magnetic field in the vicinity of non-relativistic collisionless shocks in SNRs and gamma-ray bursts (GRBs) afterglows is amplified to values greatly exceeding the ambient magnetic field both in the upstream and downstream. In the past few years, high resolution X-ray observations have provided indirect evidence for magnetic fields of values as high as $100\muG$ in the ($\textrm{v}_S\sim\mbox{few }\times1000\km\se^{-1}$) shocks of young SNRs \citep[see][]{Bamba03,Vink03,Volk05}. For relativistic shocks in the afterglow phase of GRBs, a very large magnetic field in the downstream is inferred from measurements \citep[see][for review]{Waxman06}, that can not be explained as a mere compression of the ambient field and requires a fair fraction of the kinetic energy of the shock to be transformed into magnetic energy. \citet{Li06} have furthermore shown that in GRB afterglow shocks strong amplifications of the magnetic field in the upstream can be also inferred. We therefore assume that the magnetic field is amplified by the shock both in the upstream and in the downstream to values close to equipartition
\begin{equation}\label{eq:magneticfield}B_d=\sqrt{\ep_B 8\pi \rho \vv_S^2 \gamma_S^2}.\end{equation}

Initially, the shock and ejecta velocities (which are similar) are constant with time while the radius increases linearly. For the density distributions considered here the maximal acceleration energy is attained when the ejecta begins to decelerate. This happens roughly when the energy in the swept up material is comparable to the initial ejecta energy, i.e. when $M_{\rm swept}\sim M_{\rm ej}/\gamma_{\rm ej}$. In case the shock is propagating into a homogeneous ISM, the deceleration radius is given by
\begin{equation}\label{eq:rdechom} r_{\rm hom}\approx \left(\frac{M_{\rm ej}}{\frac{4\pi}{3}n_0 m_p \gamma_{\rm ej}}\right)^{1/3},\end{equation}
where $n_{0}$ is the ISM particle density per $\mathrm{cm^{3}}$. Using Eqs. (\ref{eq:maxegeneral}), (\ref{eq:magneticfield}) and (\ref{eq:rdechom}), the maximum energy of an accelerated particle is
\begin{equation}\label{eq:EISM}
\vep<3\X10^{17}Z_{1}\epsilon_{B,-1}^{1/2}\left(\frac{M_{{\rm ej}}}{10 M_{\odot}}\right)^{1/3}n_{0}^{1/6}\beta_{\rm ej,-2}^{2}\gamma_{\rm ej}^{2/3}\,{\rm eV}.\end{equation}
Here $\epsilon_{B}=0.1\epsilon_{B,-1}$ and $\beta_{{\rm ej}}=10^{-2}\beta_{{\rm ej,-2}}$.

For a shock propagating into a progenitor wind with a density profile $\rho=\dot{M}/4\pi r^2 \vv_w$, where $\dot{M}$ is the mass loss rate and $\vv_w$ is the wind velocity, the deceleration radius is
\begin{equation}\label{eq:rdecwind}
r_{\rm wind}\approx \frac{M_{\rm ej}\vv_{w}}{\gamma_{\rm ej}\dot M}.
\end{equation}
Using Eqs. (\ref{eq:maxegeneral}), (\ref{eq:magneticfield}) and (\ref{eq:rdecwind}), the maximum energy of an accelerated particle is
\begin{equation}\label{eq:Ew}
    \vep<10^{16}Z_1\epsilon_{B,-1}^{1/2}
    \left(\frac{\dot{M}_{-5}}{\vv_{w,8}}\right)^{1/2}\beta_{\rm ej,-2}^2\gamma_{\rm ej}\,{\rm eV}.
\end{equation}
Here $\dot{M}=10^{-5}\dot{M}_{-5}\mathrm{M_{\odot}/yr}$ and $\vv_{w}=10^{8}\vv_{w,8}\mathrm{cm/s}$. The maximum energy is independent in this case of the ejecta mass, i.e. of the deceleration radius, since for a $1/r^2$ density profile the maximum energy (for a fixed shock velocity) is independent of the shock radius.

Note, that this estimate is valid provided shock deceleration occurs while it is propagating within the $\rho\propto r^{-2}$ wind profile. The $\rho\propto r^{-2}$ dependence of wind density extends up to a radius where  $\rho_w v_w^2\approx P_{ISM}$, and the wind mass contained within this radius is
\begin{equation}\label{eq:Mwestimate}
M_w\approx \frac{\dot M ^{3/2}}{(4\pi v_w P_{ISM})^{1/2}}\approx 0.3M_\odot \frac{\dot M_{-5}^{3/2}}{v_{w,8}^{1/2}P_{ISM,-0.5}^{1/2}},
\end{equation}
where the ISM pressure is $P_{ISM}=0.3\mathrm{eV/cm^3}P_{ISM,-0.5}$. The maximum energy is given by eq.~(\ref{eq:Ew}) for $M_{\rm ej}/\gamma_{\rm ej}<M_{w}$, and it is lower otherwise.

\subsection{Energy distribution within the ejecta}
\label{sec:ejecta}

SN explosions produce non-uniform ejecta, with velocity rising towards the front edge of the ejecta. The shock driven by the fastest, outermost part of the ejecta is capable of accelerating particles to the highest energy.  Here we consider the case where a small and fast part of the ejecta is responsible for accelerating cosmic rays to energies $\vep_h\sim 10^{18}\eV$, and constrain the energy distribution of the slower parts of the ejecta by requiring that they do not produce a flux of lower energy CRs that exceeds the observed one.

We denote the amount of energy in ejecta moving faster than $\bt c$ by $E_k(>\bt)$. Since we consider in this section a limited range of particle energies, we assume that each part of the ejecta contributes a constant fraction $\zt_{\rm ej}$ to each logarithmic bin of particle energies at $\vep<\vep_{\rm max}(\bt)$ (ignoring the possible energy dependence of $\zt_{\rm ej}$). Under this assumption, the resulting energy flux per logarithmic particle energy interval is $\vep^2n(\vep)\propto \tau_{c}(\vep)E_k(>\bt(\vep))$. Here $\bt(\vep)$ ($\propto \vep^{1/2}$ for $\bt\ll1$) is the shock velocity required for accelerating up to $\vep$, i.e. $\vep_{\max}(\bt)=\vep$.

Assuming that the part of the ejecta with velocities larger than some $\bt_h$ is responsible for the CR spectrum at some $\vep_h$, requiring the slower parts of the ejecta do not produce a flux of cosmic rays at lower energy that exceeds the observed flux implies
\begin{equation} \label{eq:constraint_spect1}
\frac{E_k(>\bt(\vep))}{E_k(>\bt(\vep_{h}))}<
\frac{\vep^2n_{\rm obs}(\vep)}{\vep_h^2n_{\rm obs}(\vep_h)}
\left[\frac{\tau_{c}(\vep)}{\tau_{c}(\vep_{h})}\right]^{-1},
 \end{equation}
where $n_{\rm obs}$ stands for the observed CR number density (per unit CR energy).
Since the measured energy flux per logarithmic particle energy interval in the range $10^{15}\mathrm{eV}<\vep<10^{18}\eV$ is roughly proportional to $E^{-1}$, and the confinement time is expected to decrease with energy, the energy distribution must satisfy
\begin{equation} \label{eq:constraint_spect2}
\frac{E_k(>\bt(\vep))}{E_k(>\bt(\vep_{h}))}<\left(\frac{\vep}{\vep_h}\right)^{-1}.
 \end{equation}

\section{Application to SNe and TRSNe}
\label{sec:Application}

\subsection{SNe}
\label{sec:SNe}

Eqs. (\ref{eq:EISM}) and (\ref{eq:Ew}) imply that typical SNe driven shocks,
characterized by $\bt_{ej}\sim 10^{-2}$, are probably too slow to accelerate
particles to $10^{18}\eV$, especially considering that these are optimistic
upper limits. Another difficulty in attributing the production of
$10^{18}\eV$ CRs to ordinary SNe is related to the constraint of
Eq. (\ref{eq:energeticconst2}). According to this constraint, if SNe were
capable of accelerating particles to energies as high as $10^{18}\eV$, the flux produced
at these energies would exceed the observed flux by a large factor unless
$\zeta\sim10^{-4}$, i.e. unless the energy carried by $10^{18}$~eV CRs
accelerated by the SN shock is smaller by a factor of $10^4$ than that
carried by $10^{9}$~eV CRs accelerated by the same shock. Such a strong
suppression is not expected for the case when the maximum acceleration energy
exceeds $10^{18}$~eV, since in this case we expect equal energy to be
deposited per logarithmic CR energy interval (see discussion at the end of section \S\ref{sec:Rates}). For example, attributing this factor to an acceleration spectrum $dn/d\vep\propto\vep^{-p}$ with $p$ different from $2$, would require $p>2.4$ which seems inconsistent with theory and SNR radio observations.

One may argue that the above problem may be circumvented by assuming that $10^{18}\eV$ CRs are produced only by the fast, $\bt\gg 10^{-2}$, component of the SN ejecta, which carries only a small fraction of the total kinetic energy. This would lead, however, to violation of the constraint of Eq.~(\ref{eq:constraint_spect2}). The typical energy distribution in SN ejecta is $E_k(>\bt)\propto \bt^{-5}$ for BSG progenitors and $E_k(>\bt)\propto \beta^{-6}$ for RSG progenitors \citep{Matzner99}. At non relativistic velocities $\vep_{\rm max}\propto \beta^2$, implying $E_k(>\bt(\vep))\propto \vep^{-5/2}$ or $E_k(>\bt(\vep))\propto \vep^{-3}$ for BSGs and RGBs respectively. This is in clear contradiction with Eq.~(\ref{eq:constraint_spect2}). That is, if one assumes that the fast part of the ejecta produces the observed $10^{18}\eV$ CR flux, then the predicted flux at lower energies would far exceed the observed flux.

\citet{BellLucek} have argued that SN driven shocks are candidate sources of
CRs above $10^{18}\eV$, based on an estimate of the maximum acceleration
energy. They have considered a shock of velocity of $40,000\mathrm{km/s}$
driven into a massive wind, characterized by $\dot{M}_{-5}\sim1$ and
$\vv_{w,8}\sim10^{-2}$, as inferred from radio observations of
SN1993J. Indeed, for such parameters the SN driven shock may allow
acceleration to $>10^{18}\eV$, as indicated by Eq.~(\ref{eq:Ew}). However, it
should be realized that SN1993J is still young, and that the measured
velocity therefore represents only the fast front edge of the ejecta, which
contains only a small fraction of the ejecta mass and energy. This is well
illustrated in the modeling of this SN by \citet{Woosley94}, where only a
small fraction of the total mass and energy lies in such a high velocity component of the ejecta (see e.g. their fig. 6). As explained above, if the tail of the ejecta velocity distribution can account for the cosmic ray flux at some energy $\vep_h\sim10^{18}$~eV, then the steep profile of energy as a function of velocity would imply a CR flux at lower energies that far exceeds the observed flux.

\subsection{TRSNe}
\label{sec:TRSNe}

During the past decade, four GRBs were observed to be associated with core collapse SNe of type Ic: GRB980425/SN1998bw \citep{Galama98}, GRB030329/SN2003dh \citep{Hjorth03,Stanek03}, GRB031203/SN2003lw \citep{Tagliaferri04}, and XRF060218/SN2006aj \citep{Campana06}. All but GRB030329 were very faint, 4-5 orders of magnitude fainter than an average cosmological GRB. In all 4 cases there is evidence from radio observations for deposition of a significant part of the kinetic energy in a mildly relativistic part of the ejecta \citep[][and references therein]{Soderberg06}. Although radio observations alone do not allow one to uniquely determine the energy carried by the relativistic ejecta \citep{WL99,Waxman04a}, additional observational information enabled such a unique determination in two cases. For GRB980425/SN1998bw there is a robust estimate of the energy deposited in the fast, $\beta\ge0.85$, part of the ejecta, based on long term radio and X-ray observations, which yield $E(\beta\ge0.85)=10^{49.7}$~erg \citep{Waxman04b}. Similar values are inferred for XRF060218/SN2006aj from prompt and afterglow X-ray observations \citep{Waxman07}.

The SN light curves of SN1998bw, SN2003dh, and SN2003lw appear to be qualitatively similar, implying SN ejecta mass and energy of $E_{ej}\sim5\times10^{52}$~erg, $M_{ej}\sim10M_{\odot}$ \citep[e.g. table 6 of][]{Mazzali06a}. Note, that the energy may be lower, $\sim1\times10^{52}$~erg, if the explosion is not spherical, which is likely \citep{Maeda06}. The SN light curve of SN2006aj is qualitatively different than the others, implying $E_{ej}\simeq2\times10^{51}$~erg and $M_{ej}\simeq2M_\odot$ \citep[in this case, the explosion is probably not highly non spherical;][]{Mazzali06b}. The inferred $E_{ej}$ and $E(\beta\ge0.85)$ imply that a significant fraction of the kinetic energy is deposited in a fast, mildly relativistic part of the ejecta, $E(\beta\ge0.85)/E_{ej}\ge10^{-2}$.

Two points should be emphasized here. First, GRBs associated with TRSNe are
different than cosmological GRBs: The amount of energy carried
by relativistic ejecta in TRSN explosions, $\sim10^{49.5}$~erg, is much lower
than that of cosmological GRBs, which is $\sim10^{51}$~erg, and
the fastest part of the ejecta in a TRSN is only mildly relativistic,
unlike the highly relativistic ejecta inferred from observations of GRBs.
Second, the
mechanism responsible for depositing $\sim10^{49.5}$~erg in the mildly
relativistic part of the ejecta is not understood. For the inferred values of
$E_{ej}/M_{ej}c^2\sim10^{-3}$, the acceleration of the SN shock near the edge
of the star is expected to deposit only $\sim10^{-7}E_{ej}$ in the part of
the ejecta expanding with $\gamma\beta>1$ \citep{Tan01}. This suggests that
the mildly relativistic component is driven not (only) by the spherical SN
shock propagating through the envelope, but possibly by a more relativistic
component of the the explosion, e.g. a relativistic jet propagating through
the star \citep{Aloy00,Zhang03}. For the present analysis, we use therefore only the observational constraints on the energy distribution within the ejecta.

With only 4 detected events, the TRSNe rate is rather uncertain. Estimates of the local ($z=0$) rate range from $\sim10^2$ to $\sim10^{3}\mathrm{Gpc}^{-3}\yr^{-1}$ \citep{Guetta07,Soderberg06,Liang06,Pian2006}. Comparing this to the local SN rate, $\approx10^{5}\mathrm{Gpc}^{-3}\yr^{-1}$ \citep[e.g.][]{Cappellaro99}, we find $\dot{N}_{\rm TRSN}/\dot{N}_{\rm SN}\approx10^{-2.5\pm0.5}$. This estimate is consistent with the estimate of \citet{Podsiadlowski04} of the rate of "hypernovae" in galaxies similar to the Milky Way. As mentioned in the introduction, the TRSNe rate may be higher if some TRSNe are not associated with GRBs (and hence were not identified following a GRB trigger). With these estimates at hand, we are now ready to determine whether or not TRSNe meet the constraints derived in \S~\ref{sec:physical_constraints}.

First, for $\bt\approx 0.8$ Eqs.~(\ref{eq:EISM}) and~(\ref{eq:Ew}) imply that
the mildly relativistic part of TRSNe ejecta may indeed accelerate CRs up to
energy exceeding $10^{18}\eV$. Second, for an energy $E_r\sim10^{49.5}$~erg
in the relativistic part of the ejecta we have $E_{\rm s}/E_{\rm SN}\sim10^{-1.5}$, satisfying the energetics constraint, Eq.~(\ref{eq:jointconst}), and $\dot{N}_{\rm TRSN}/\dot{N}_{\rm SN}\approx10^{-2.5\pm0.5}$ satisfying the rate constraint, Eq.~(\ref{eq:energeticconst2}). Note, that for TRSNe we expect $\zeta\sim1$, based on their radio and X-ray observations. These observations imply that the collisionless shocks driven by the mildly relativistic ejecta into the wind surrounding the progenitor transfer a significant fraction of the energy to a population of shock accelerated electrons, and that the accelerated electron spectrum follows $\vep^2 dN/d\vep\approx\vep^0$ over a wide range of energies, from electrons emitting synchrotron radiation in the radio band over at least $\sim5$ decades of energy to those emitting radiation in the X-ray band \citep{Waxman04b}. Thus, if nuclei are accelerated with similar efficiency to a similar energy power spectrum, the fraction of energy deposited in $\sim10^{18}$~eV CRs would be similar to that deposited by SNe in CRs accelerated to $\sim10^9$~eV.

Let us next consider the constraint of Eq.~(\ref{eq:constraint_spect1}) on the energy distribution within the ejecta. Consider a TRSN with a total kinetic energy of $E_{ej}=10^{52}E_{ej,52}\erg$, an ejecta mass $M_{\rm ej}=10M_{\rm ej,1}M_{\odot}$ and an energy $E_r$ in a mildly relativistic, $\bt\ge0.8$, part of the ejecta. Using Eq.~(\ref{eq:Ew}), the ratio of the maximum energies of particles accelerated by the relativistic part of the ejecta, $\vep_{\max,r}$, and by the bulk of the ejecta, $\vep_{\max,ej}$, is
\begin{equation}
\vep_{\max,r}/\vep_{\max,ej}\approx 10^3 E_{\rm ej,52}^{-1}M_{\rm ej,1},
\end{equation}
and the constraint of Eq.~(\ref{eq:constraint_spect1}) is satisfied in the energy range of $10^{15}\eV-10^{19}\eV$ as long as
\begin{equation}
E_{r}>10^{49} \frac{E_{\rm ej,52}^2}{M_{\rm ej,1}}\frac{\tau_{c}(\vep_{\max,r})}{\tau_{c}(\vep_{\max\rm,ej})}\,{\rm erg}.
\end{equation}
Assuming that $\tau_{c}\propto \vep^{\de}$ we have
\begin{equation}\label{eq:Er_min}
E_{r}>10^{49} \frac{E_{\rm ej,52}^{2-\de}}{M_{\rm ej,1}^{1-\de}}10^{3\de}\,{\rm erg}.
\end{equation}

An upper limit to the value of $\delta$ in the relevant energy range, $10^{15}\eV-10^{19}\eV$, may be obtained as follows. First note that the confinement time of $Z=10$ particles at $\vep=10^{9}Z\eV$ is $10^{7.5}\yr$ while and at $\vep=10^{17}Z\eV$ it is larger than $10^{4.5}\yr$, implying that the average value of $\de$ between these energies satisfies \[\bar{\de}\equiv \frac{\log(\vep_h/\vep_l)}{\log[\tau_{c}(\vep_h)/\tau_{c}(\vep_l)]}\leq 0.4\,.\] In fact, the value of $\delta$ in the energy range of $10^{15}\eV-10^{19}\eV$ must be lower than this average. The grammage (column density) traversed by CRs of energy $<1\TeV$ before they escape our Galaxy is $\Sigma_{\rm conf}\approx 9(E/10Z\,\GeV)^{-\de}\rm{g} \cm^{-2}$ with $\de\approx0.6$ \citep{Engelmann90,Stephens98,Webber03},
suggesting $\tau_{c}\propto 10^{7.5}(E/Z\,\GeV)^{-0.6}\yr$ for $E/Z<10^{12}\eV$. As the value of $\de$ at energies $10^9Z\eV-10^{12}Z\eV$ is higher than the average, $\bar\de\le0.4$, the value of $\de$ must be lower at energies $10^{13}Z\eV-10^{17}Z\eV$. It is reasonable to assume therefore $10^{3\de}<10$, which implies that the constraint of Eq.~(\ref{eq:Er_min}) may be satisfied for reasonable values of the parameters.

Finally, we note that \citet{Milgrom96} have suggested that cosmological GRBs, which occur in our galaxy once every $\sim10^5$ years, could be the sources of the CRs in the energy range of $10^{14}\eV$ to $10^{19}\eV$. This suggestion is quite distinct from the one presented in this article. While \citet{Milgrom96} considered acceleration by the highly relativistic jets of cosmological GRBs, where $\sim10^{51}$~erg is carried by a highly relativistic, $\gamma\sim10^{2.5}$ ejecta, we consider acceleration by mildly relativistic, $\gamma\beta\sim1$, less energetic, $\sim10^{49.5}$~erg, and more frequent TRSN explosions.

\section{Discussion}
\label{sec:conclusions}

We derived constraints that must be satisfied by the sources of $\sim10^{15}$ to $\sim10^{18}$~eV cosmic rays, under the
assumption that the sources are Galactic and that the CRs in this energy range are confined by the Galactic magnetic field (as explained in \S~\ref{sec:Rates},
the CRs in this energy range are likely to be confined by the Galactic magnetic field since their Larmor radius is much
smaller than the CR disk thickness).
In \S~\ref{sec:Rates} we have shown that the Galactic occurrence rate, $\dot N_{\rm s}$, of sources producing
$\sim10^{18}$~eV CRs and their energy production $E_{\rm s}$ per event must satisfy the constraints given by
Eqs.~(\ref{eq:jointconst}) and~(\ref{eq:energeticconst2}) [$E_{\rm s}$ is the energy available for the acceleration of $10^{18}\eV$ CRs, rather than the total event energy, as explained preceding eq.~(\ref{eq:sourcespectrum})].
In particular, $E_{\rm s}$ must be considerably lower than the kinetic energy of a typical SNe,
$E_{\rm s}<0.03\dot N_{SN,-2}\zeta^{-1}E_{\rm SN}$, where $\dot N_{\rm SN}=0.01\dot N_{SN,-2}\yr^{-1}$ is the Galactic SNe rate.
$\zeta$ in these equations is the ratio of the fraction of $E_{\rm s}$ deposited in $10^{18}$~eV CRs to the fraction of the kinetic SN energy that is deposited in $\sim10^9$~eV CRs. These
constraints must be satisfied by any candidate Galactic source. In \S~\ref{subsec:Maximum-energy} and \S~\ref{sec:ejecta}
we derived additional constraints that are applicable to SNe and TRSNe. These sources are expected to accelerate
particles to high energies through the collisionless shocks that they drive into the plasma surrounding the exploding
star. In \S~\ref{subsec:Maximum-energy} we derived constraints on the velocity of the shells ejected by the explosions [eqs.~(\ref{eq:EISM}) \&~(\ref{eq:Ew})],
and in \S~\ref{sec:ejecta} we derived constraints on the structure of the ejecta [eqs.~(\ref{eq:constraint_spect1}) \&~(\ref{eq:constraint_spect2})].

In \S~\ref{sec:SNe} we have shown that ordinary SNe are unlikely to be the sources of $\sim10^{18}$~eV CRs, since they are unlikely to be able to accelerate particles to such energy and if they could, they would produce fluxes far exceeding the observed flux at these energies. We have furthermore shown that assuming that the flux of $\sim 10^{18}\eV$ CRs is produced by the fastest part of the ejecta of SNe, which carry a small fraction of the ejecta energy, would predict a CR flux at lower energy, $\sim10^{15}$~eV, which far exceeds the observed flux.

In \S~\ref{sec:TRSNe} we have shown that Galactic TRSNe may be the sources of high energy, $\sim10^{18}\eV$, CRs.
The mildly relativistic shocks driven by such ejecta are likely to be able to
accelerate particles to energies exceeding $10^{18}\eV$.  The estimated rates
of TRSNe combined with the typical energy they deposit in mildly relativistic
ejecta yield a flux of $10^{18}\eV$ CRs which is comparable to the observed
flux (under the assumption that the efficiency of the acceleration of
particles in these shocks is similar to that of SNe and that the accelerated
particle spectrum is close to $dN/d\vep\propto\vep^{-2}$, which, as explained
in \S~\ref{sec:TRSNe} is supported by radio and X-ray observations).  The
large fraction, $\ge10^{-2}$, of the kinetic energy deposited by such
explosions in mildly relativistic ejecta ensures that if TRSNe are
responsible for producing the observed flux at $10^{18}\eV$ they do not
overproduce lower energy CRs.  Furthermore, since the TRSNe are a subset of
the larger SNe population, their CR production is expected to smoothly join
the lower energy CRs produced by SNe.

\acknowledgments This research was partially supported by AEC, Minerva and
ISF grants (RB, BK, EW). EW and BK thank the Institute for Advanced Study for
its hospitality during a period where part of this research was carried out.
AM acknowledges support from the Frank and Peggy Taplin membership at the
Institute for Advanced Study.  While this paper was in preparation, we became
aware of work in preparation by X. Wang, S. Razzaque, P. Meszaros and Z. Dai
proposing hypernovae as cosmological sources of high energy CRs \citep{Wang07}.
In the present work we have focused on Galactic sources.

\bibliographystyle{apj}

\end{document}